\documentclass[pdflatex,sn-mathphys-num]{sn-jnl}
\usepackage{nicematrix}
\usepackage{graphicx}%
\usepackage{multirow}%
\usepackage{upgreek}
\usepackage{amsmath,amssymb,amsfonts}%
\usepackage{amsthm}%
\usepackage{braket}
\usepackage{mathrsfs}%
\usepackage[title]{appendix}%
\usepackage{xcolor}%
\usepackage{textcomp}%
\usepackage{manyfoot}%
\usepackage{booktabs}%
\usepackage{algorithm}%
\usepackage{algorithmicx}%
\usepackage{algpseudocode}%
\usepackage{tabularray}
\UseTblrLibrary{siunitx}
\usepackage{listings}%
\newcommand{\ketbra}[2]{\ket{#1}\bra{#2}}
\usepackage{subcaption} 
\begin{document}
\title[A reconfigurable entanglement distribution network suitable for connecting multiple ground nodes with a satellite]{A reconfigurable entanglement distribution network suitable for connecting multiple ground nodes with a satellite}
\author*[1]{\fnm{St\'ephane} \sur{Vinet}}\email{svinet@uwaterloo.ca}

\author[1,2]{\fnm{Ramy} \sur{Tannous}}\email{ramy.tannous@uwaterloo.ca}

\author[1,3]{\fnm{Thomas} \sur{Jennewein}}\email{thomas.jennewein@uwaterloo.ca}

\affil*[1]{\orgdiv{Institute for Quantum Computing and Department of Physics \& Astronomy}, \orgname{University of Waterloo}, \orgaddress{\street{200 University Ave W}, \city{Waterloo}, \postcode{N2L 3G1}, \state{Ontario}, \country{Canada}}}

\affil[2]{\orgdiv{Currently with the National Research Council of Canada}, \orgaddress{\street{100 Sussex Drive}, \city{Ottawa}, \postcode{K1A 0R6}, \state{Ontario}, \country{Canada}}}

\affil[3]{\orgdiv{Department of Physics}, \orgname{Simon Fraser University}, \orgaddress{\street{8888 University Dr W}, \city{ Burnaby}, \postcode{V5A 1S6}, \state{British Columbia}, \country{Canada}}}


\abstract{Satellite-based quantum communication channels are important for ultra-long distances. Given the short duration of a satellite pass, it can be challenging to efficiently connect multiple users of a city-wide network while the satellite is passing over that area. We propose a network with dual-functionality: during a brief satellite pass, the ground network is configured as a multipoint-to-point topology where all ground nodes establish entanglement with a satellite receiver. During times when this satellite is not available, the satellite up-link is rerouted via a single optical switch  to the ground nodes, and the network is configured as a pair-wise ground network. We numerically simulate a pulsed hyper-entangled photon source and study the performance of the proposed network configurations for quantum key distribution. We find favourable scaling in the case that the satellite receiver exploits time-multiplexing whereas the ground nodes utilize frequency-multiplexing. The scalability, simple reconfigurability, and easy integration with fibre networks make this architecture a promising candidate for quantum communication of many ground nodes and a satellite, thus paving the way towards interconnection of ground nodes at a global scale.}


\maketitle

\section{Introduction}
A primary objective of quantum communication is the distribution of entanglement between distant parties \cite{Wehner}. This serves as the foundation for creating the quantum internet: a quantum network that enables revolutionary applications, including quantum key distribution (QKD) \cite{bennet1984proceedings,E91}, clock synchronization \cite{komarquantum2014,ilookekeremote2018}, distributed quantum computing \cite{Broadbent,Nickerson}, and long-baseline interferometry \cite{Gottesman}. 
 In recent years, significant research efforts have been done in 
 expanding the scalability of quantum networks beyond two communicating parties. 
 To this extent, multiple network architectures have been proposed, notably trusted-node networks which consist of a mesh of point-to-point links, where each node acts as a trusted relay to establish a complete two-party communication set-up \cite{wengerowskyentanglementbased2018,Stucki,Tokyo,2009NJPh...11g5001P,xufield2009}. While such a scheme is compatible with heterogeneous quantum links and can be used to increase the communication distance arbitrarily, it comes with a significant security cost as one must ensure the privacy of every relay site. Alternatively, active switch quantum networks \cite{2003switch,NIST,darpa,Chen:10,chenintegrated2021} allow for dynamically reconfigurable networks, yet are vulnerable to device failure. Furthermore, the distribution rate is limited by the switch's operating speed. Hence, an all-passive solution is preferred where all users can be connected simultaneously. 
 
Two main approaches have previously been considered to create fully-connected quantum networks relying either on high-dimensional entanglement or bi-partite entanglement and multiplexing \cite{1999AIPC..461..220T,wengerowskyentanglementbased2018,2016LPRv...10..451A,2008OExpr..1616052L,Pseiner2021,Ecker,krzic_towards_2023}. In the multipartite method, each user pair shares a different subspace of the overall network Hilbert space, which is used to generate their keys. However, adding or subtracting users from the network requires changing the dimensionality of the state produced by the source making this approach unscalable. On the other hand, multiplexing exploits the inherent hypercorrelations produced by the entangled photon source to deterministically separate correlated photon pairs to different users. 
 This architecture is easily scalable as the number of users in the network can be modified without any change to the photon source. Nevertheless for an $n$-node network to be fully connected, $\mathcal{O}(n^2)$ multiplexed channels are required. Consequently, there remains an important hardware requirement as each channel requires a decoder and detector module to be fully distinguishable. Otherwise an increase in accidental counts will deteriorate the performance of the network as was shown in \cite{wengerowskyentanglementbased2018, Pseiner2021}.
 Recently, \cite{Joshi,qi15user2021,liu40user2022,Liu:23} utilized beamsplitters in addition to multiplexing to increase the scalability of their network, but at the expense of the key generation rate. An interesting solution proposed by \cite{wengerowskyentanglementbased2018}, is to use a pulsed source to recover the channel distinguishability despite limited detector resources. This implementation requires an additional gating signal to be sent to each user, and results in noise reduction by a factor equal to the gating duty cycle. However, this scheme is unsuitable for mobile nodes as the delays to every other node would need to be continuously compensated by each user thus restricting the reach of these quantum networks to metropolitan areas. Continued extensive research is being conducted on fibre-based quantum repeaters \cite{2019NaPho,2021NaPho..15..374P,2021PhR} and satellite-based quantum links \cite{2017Natur.549...43L,PhysRevLett.119.200501,2018Liao,2020Yin} to enable the establishment of a global-scale network. While fibre-based links are limited in distance by exponential losses, the use of a free-space link to an orbiting satellite reduces the loss scaling from exponential to quadratic, consequently extending the reach of individual quantum links \cite{Bourgoin_2013}.  

 In this paper we present a scalable quantum network architecture that includes a mobile satellite node with minimal hardware requirements. To interface the fibre-optical network with the satellite link, we consider a pulsed non-degenerate entangled photon source in conjunction with a frequency-to-time mapping \cite{Davis:17,FBGarray}. Using a medium with a high group delay dispersion (GDD), one can carefully map the frequency distribution of pulsed light onto temporal modes such that each frequency channel obtains a distinct time delay within the resolution of the detector. The satellite node is then able to unambiguously separate the channels while using a single detector module.
  The rest of the paper is organized as follows. In Section \ref{Method}, we introduce the reconfigurable quantum network architecture. In Section \ref{results} we analyze the performance of the proposed network configuration and benchmark its potential by studying its use case for the Canadian QEYSSat quantum satellite mission \cite{qeyssat}. A summary of our results and various extensions are given in Section 
\ref{discussion}.
\section{Method}\label{Method}
\begin{figure}[h]
\includegraphics[width=\textwidth]{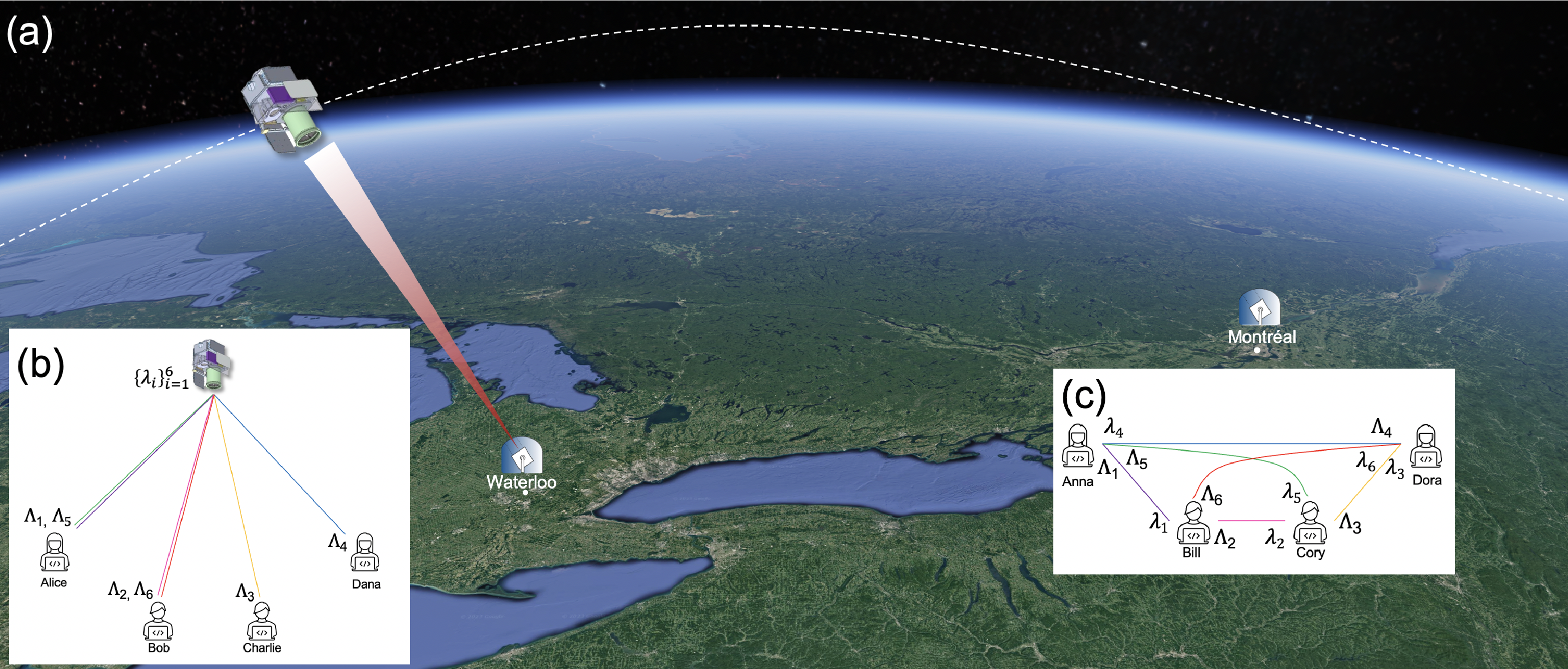}

\caption{(a) Communication layer and wavelength allocation for a satellite pass over two ground stations e.g. over Waterloo and Montreal. (b) Waterloo is in view of the satellite and takes the multipoint-to-point satellite pass configuration whereas Montreal is in a fully connected ground topology (c). At a later time, with the satellite now in view of Montreal, the two cities will adopt the opposite topology by redistributing the frequency-time correlated pairs $\{\lambda_{i},\Lambda_{i}\}$  between the nodes. During the satellite pass Alice and Bob enjoy are connected by two frequency-time channels and thus enjoy a higher key rate.  }
    \label{fig: alloc}
\end{figure}
\begin{figure}
       \includegraphics[width=\textwidth]{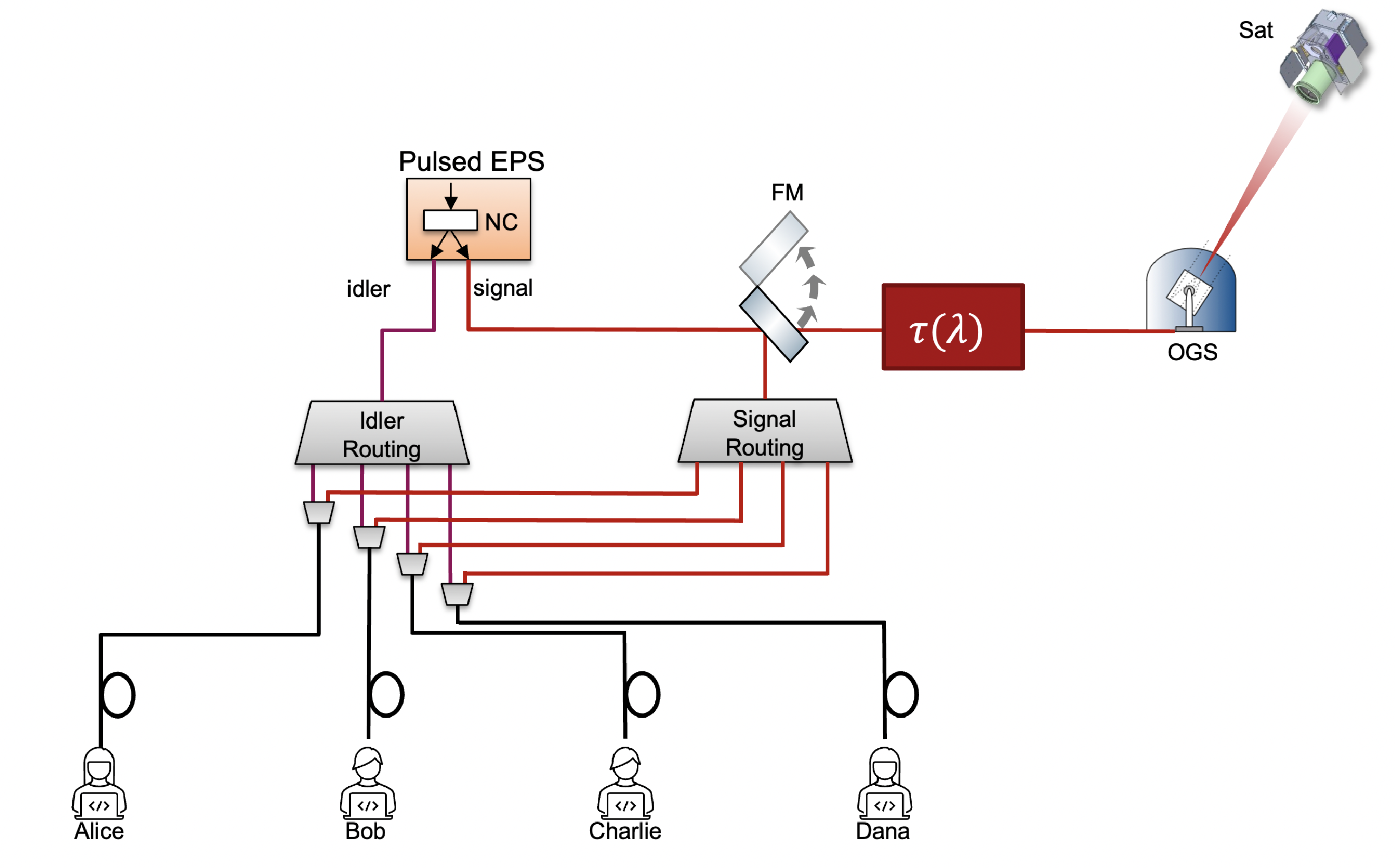}
    \caption{Conceptual physical layout of the hybrid quantum network. The entangled photon source (EPS) consists of a non-linear crystal (NC) pumped by a pulsed laser to create entangled photon pairs via spontaneous parametric down-conversion. 
    During the satellite pass (flip mirror (FM) up) the signal is sent through a frequency-to-time encoder $\uptau(\lambda)$ in an up-link to the satellite (Sat) via the optical ground station (OGS). Outside of the satellite pass (flip mirror down) the signal is rerouted with the idler through telecom fibre to the various ground nodes thus enabling a fully-connected topology.}
        \label{fig:design2}
\end{figure}

The proposed network architecture  is depicted in Fig.~\ref{fig: alloc}. It allows two reconfigurable quantum metropolitan-area networks (QMAN), for example in Waterloo and Montreal, Canada, to be connected by ground-to-satellite quantum links. As the satellite passes over Waterloo in Fig.~\ref{fig: alloc}b, the network adopts a multipoint-to-point topology from the $N$ ground nodes to the satellite. Consequently allowing for a substantive increase of the key generation during the satellite pass via multiplexing. Meanwhile, the Montreal network in Fig.~\ref{fig: alloc}c is arranged into a pair-wise configuration where every user is connected to each other. As the satellite travels out of reach of Waterloo and within reach of Montreal, each network reconfigures itself into the opposite topology to optimize its functionality throughout the day. 
This rearrangement is enabled by the hybrid quantum network design in Fig~\ref{fig:design2}. For a satellite up-link, the idler is kept on the ground and distributed to the network users. During the satellite pass, the signal is sent through a frequency-to-time encoder $\uptau(\lambda)$ to assign a time delay to each wavelength channel. This mapping can be accomplished by wavelength-dependent group delay dispersion e.g. in a chirped fibre Bragg grating \cite{Davis:17,Karpinski,Yang:20} (CFBG) or by using a fibre Bragg grating (FBG) array with suitable spacing between each FBG to obtain the desired time delay \cite{FBGarray}. Outside of the satellite pass, the flip mirror redirects the signal to the ground nodes. 

The minimal duration of the channel time separation for the frequency-to-time mapping is determined by the timing jitter of the satellite single-photon detector and readout electronics which is typically on the order of $\sim 130$ ps for Si-APDs \cite{spadjitter,GERSBACH2009803,6044205,4303048}. Thus for a 80 MHz pump repetition rate, the satellite network configuration is limited to a maximum of $\sim 95$ frequency-time channels within the 12.5 ns pulse train. To ensure the channel distinguishability, the GDD must then be greater than $\sim$ 325 ps/nm. For a given device length, the GDD and the CFBG's reflection bandwidth are inversely proportional: a larger dispersion requires higher reflectivities resulting in a narrower resonance and bandwidth \cite{Azana,Ouellette}. However, this limitation can be circumvented by using longer CFBGs or by using a CFBG array i.e. inscribing or splicing multiple CFBGs with adjacent reflection windows in succession \cite{Davis:17}. 

While the number of channels depends on the entangled photon source's brightness and bandwidth, for simplicity we assume four nodes in each QMAN in Fig.~\ref{fig: alloc}. To create a fully connected N-user communication network, $\binom{N}{2}$ links are needed. Consequently, in the satellite configuration shown in Fig.~\ref{fig: alloc}b the six channels are distributed to the users such that Alice and Bob each receive two frequency-time channels and therefore attain an increased secure key rate. For the pair-wise fully-connected topology in Fig.~\ref{fig: alloc}c, each node receives 3 wavelength channels to allow communication with every other node. The network connectivity and the wavelength allocation for the two different configurations: satellite (flip mirror up) and ground (flip-mirror down) are summarized in Table~\ref{tab:alloc}.
\section{Performance estimation}\label{results}
We simulated the key rate for various quantum key distribution scenarios using the approaches introduced by Ma et al. \cite{Ma} and Holloway et al. \cite{Holloway_paper}. Our entangled state is constructed from the following spontaneous parametric down-conversion (SPDC) Hamiltonian:
\begin{equation}\label{eq:H}
    H=i\chi(a_H^\dagger b_V^\dagger-a_V^\dagger b_H^\dagger)+ \rm{H.c.},
\end{equation}
where the subscripts $H$ and $V$ denote the horizontal and vertical polarizations, H.c. the Hermitian conjugate and $\chi$ is the squeezing parameter, a coupling constant proportional to the crystal nonlinearity and the amplitude of the pump beam \cite{TJsim}. Expectation values for the detector operators applied to the SPDC state are used in a squashing model to determine the two-fold coincidence rates and quantum bit error rate (QBER) \cite{NL,squashingNL}. We analyze the performance of both bucket and photon-number resolving (PNR) detectors whose positive operator-valued measure (POVM) are respectively given by \cite{kok_lovett_2010}:

\begin{equation}\label{eq:bucket}
\hat{E}^{(\text{click})}=\sum_{m=0}^\infty[1-D(0)(1-\eta)^m]\ketbra{m}{m}.
\end{equation}
and 
\begin{equation}\label{eq:PNR}
    \hat{E}^{(n)}=\sum_{m=0}^\infty\sum_{k=0}^n D(k) \binom{n-k+m}{m} \eta^{n-k}(1-\eta)^m\ketbra{n+m-k}{n+m-k},
\end{equation}
where $\eta$ is the detector efficiency and $D(k)$ the dark count distribution. For the satellite pass configuration, we assume the satellite detector to be a Si-APD \cite{pugh_airborne_2017} with a detector dead time of 1 $\mu$s whereas we assume the ground detector to be an SNSPD with a dead time of 10 ns  \cite{Migdall, eisaman_invited_2011}. 

\begin{table}[h]
    \centering
    \captionsetup{justification=centering}
    \caption{Proposed network connectivity and wavelength allocation. Wavelength correlated pairs are denoted as $\{\lambda_{i},\Lambda_{i}\}$ for signal and idler respectively. The specific wavelengths considered in Figs.~\ref{spectrum}-\ref{qeyssat} are provided in parentheses with their associated ITU channel number.}
    \begin{tabular}{|c|c|c|c|c|}
    \hline
      Network  & \multicolumn{2}{c|}{Satellite configuration}& \multicolumn{2}{c|}{Ground configuration}\\
       Connection &signal&idler  &signal&idler  \\ 
                \hline
       Alice-Satellite &$\lambda_1,\lambda_5$&$\Lambda_1,\Lambda_5$& - &-\\
      
       Bob-Satellite&$\lambda_2$, $\lambda_6$&$\Lambda_2$, $\Lambda_6$ & -&- \\
       Charlie-Satellite &$\lambda_3$&$\Lambda_3$& - &-\\
       Dana-Satellite&$\lambda_4$&$\Lambda_4$ & -&-\\
       Alice-Bob&-&- &$\lambda_1$  ($786.75$ nm) & $\Lambda_1$ (Ch40 $=1545.32$ nm) \\
        Bob-Charlie &-&-&$\lambda_2$  ($787.03$ nm)& $\Lambda_2$ (Ch41 $=1544.53$ nm)\\
        Charlie-Dana &-&-&$\lambda_3$  ($787.32$ nm) &$\Lambda_3$  (Ch42 $=1543.73$ nm)\\
         Alice-Dana &-&-&$\lambda_4 $ ($787.63$ nm)& $\Lambda_4$  (Ch43 $=1542.14$ nm)\\
        Alice-Charlie &-&-&$\lambda_5 $ ($787.93$ nm)& $\Lambda_5$ (Ch44 $=1542.14$ nm) \\
                Bob-Dana  &-&-&$\lambda_6 $ ($788.24$ nm)& $\Lambda_6$ (Ch45 $=1541.35$ nm)\\
                
                \hline
    \end{tabular}

    \label{tab:alloc}
\end{table}

\begin{figure}
    \centering
    \includegraphics[width=0.5\textwidth]{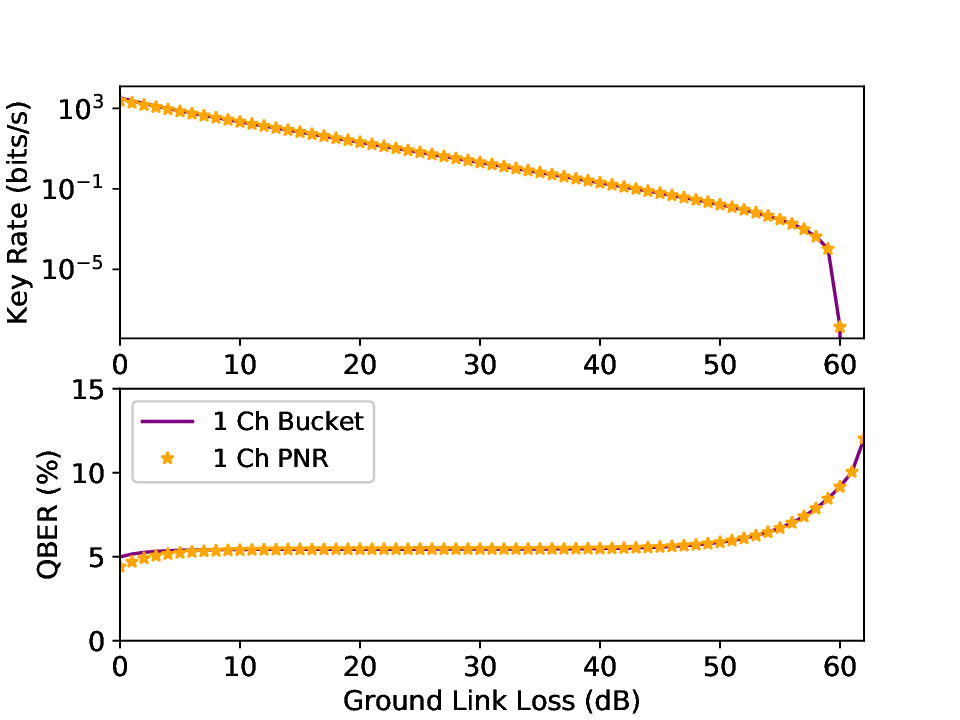}
    \caption{Secure key rate performance comparison between PNR and Bucket detectors for the ground user nodes. The simulation parameters considered can be found in Table~\ref{tab:param}.  }
    \label{fig:PNR}
\end{figure}

Due to the dead time, the photon count rate is no longer linearly proportional to the impinging photon flux but is given by: 
\begin{equation}
    N_{count}=\frac{N_{ideal}}{1+N_{ideal}T_D/T_{INT}},
\end{equation}
where $T_D$ corresponds to the dead time, $T_{INT}$ to the integration time, $N_{ideal}$ to the number of photons impinging on the detector and $N_{count}$ the number of resolved photons \cite{deadtime}. The secure key-rate (SKR) is then obtained via

\begin{equation}\label{eq:SKR}
    SKR = q\{Q_\lambda[1-f(\delta_b)H_2(\delta_b)-H_2(\delta_p)]\}
\end{equation}
where q is the basis reconciliation factor, the subscript $\lambda$ denotes for one-half of the expected photon number
$\mu$, $Q_\lambda$ is the overall gain, $\delta_b$ $(\delta_p)$ is the bit (phase) error rate, $f(x)$ is the bidirection error correction efficiency and $H_2(x)$ is the binary entropy function,
\begin{equation}
    H_2(x)=-x\log_2(x)-(1-x)\log_2(1-x).
\end{equation}
A summary of the experimental parameters assumed throughout the simulation can be found in Table~\ref{tab:param}. 
\begin{table}
    \centering
    \captionsetup{justification=centering}
    \caption{ Experimental assumptions for QKD modeling}
    \begin{tabular}{|c|c|}
    \hline
       Sat-APD dark count rate & 1000 cps \\
       Ground detector dark count rate & 100 cps\\
       Sat-APD timing jitter & 130 ps\\
       Sat-APD dead time & 1 $\mu$s\\
       Ground detector dead time & 10 ns\\
       coincidence time window & 1 ns\\
        Error correction factor & 1.17\\
        Satellite link loss & 40 dB\\
        \hline
    \end{tabular}

    \label{tab:param}
\end{table}
\begin{figure}
    \centering
        \begin{subfigure}{.495\textwidth}
      \centering
       \includegraphics[width=\textwidth]{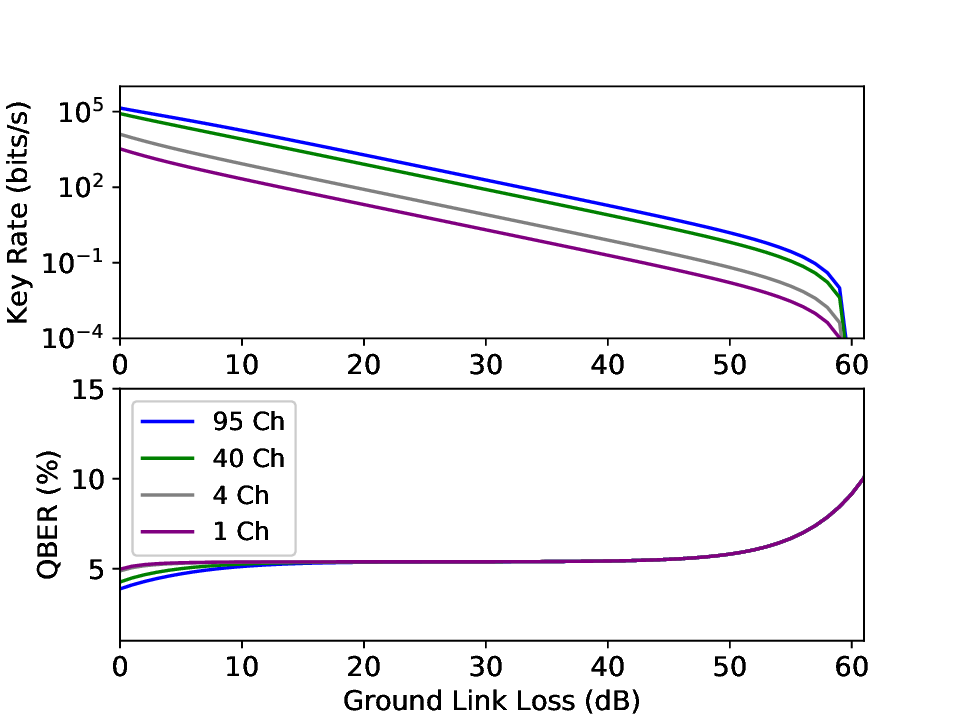}
      \caption{With time-frequency mapping}
         \label{fig:pointmultipointconfig}
          \end{subfigure}
    \begin{subfigure}{.495\textwidth}
      \centering
      \includegraphics[width=\textwidth]{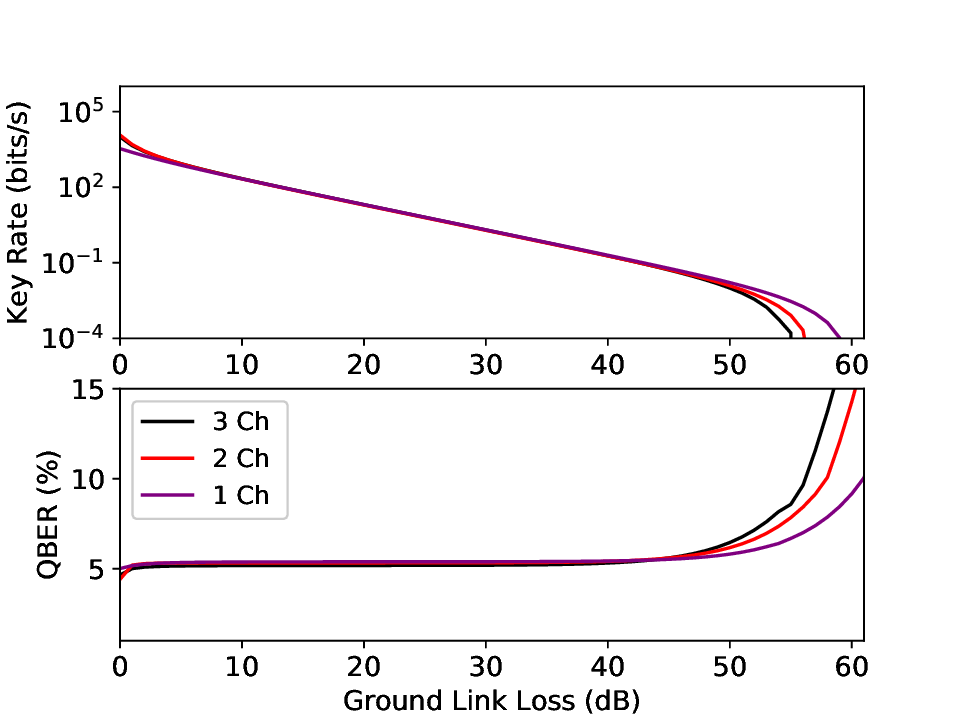}
      \caption{Without time-frequency mapping}
         \label{fig:onesided}

    \end{subfigure}%

   \caption{Scaling behaviour of the ground-to-satellite QKD scenarios for (a) with the time-frequency mapping. The time and frequency multiplexing leads to a linear key rate improvement. (b) Without the $\uptau(\lambda)$ frequency-to-time encoder, there is no discrimination made between the wavelength channels jointly recorded at the satellite resulting in an increase in accidental counts and reduction in key rate. The squeezing parameter $\chi$ in Eq.~\ref{eq:H} was optimized for every channel setting to maximize the SKR. Realistic detector parameters were considered and can be found in Table~\ref{tab:param}. }
   \label{fig:results}
\end{figure}
At low ground link loss ($<10$ dB) in Fig.~\ref{fig:pointmultipointconfig}, detector saturation at the satellite node partially limits the scaling performance. However, in the higher loss regime one obtains a linear key rate increase whilst maintaining the same QBER. This scaling stems from the deterministic separation of the multiplexed channels resulting in independent communication channels with the same signal-to-noise ratio \cite{wengerowskyentanglementbased2018}. To achieve this deterministic scenario, the frequency-to-time mapping $\uptau(\lambda)$ plays a crucial role. This is particularly apparent, in Fig.~\ref{fig:onesided} where we only consider wavelength multiplexing. In other words, the idler photons are demultiplexed on the ground but the signal photons are recorded at the satellite detector without any channel discrimination. Surprisingly, demultiplexing solely the idler at the ground stations provides little advantage and in fact results in a maximal tolerable loss decrease compared to a single channel. This is due to an increase in accidental coincidences resulting in a loss of fidelity caused by detecting several channels on the single satellite detector \cite{Pseiner2021}. Therefore, the signal-to-noise ratio decreases proportionally to the number of users added. Note that for every channel link setting, the squeezing parameter ($\chi$) was numerically optimized to maximize the SKR. To compensate for the increase in accidental counts at higher loss, the photon pair production was decreased proportionally to the number of channels considered. At low loss, this one-sided multiplexing approach yields a slight key rate increase due to detector load management bypassing the dead time of the detector. Nonetheless, for realistic loss scenarios detector saturation has negligible significance. We investigate in Fig.~\ref{fig:PNR} the use of PNR detectors at the ground user nodes. Interestingly, we find that PNR detectors offer no significant key rate advantage compared to bucket detectors.
This can be readily seen in the high loss regime from considering the series expansion of $\hat{E}^{(1)}$ for $D(1)\ll 1, \eta \to 0$ at which point one finds that $\hat{E}^{(1)}\to \hat{E}^{(\text{click})}$. A detailed analysis of this phenomenon is in preparation \cite{SV_PNR}.

\begin{figure}
\begin{subfigure}{0.49\textwidth}
    \includegraphics[width=\textwidth]{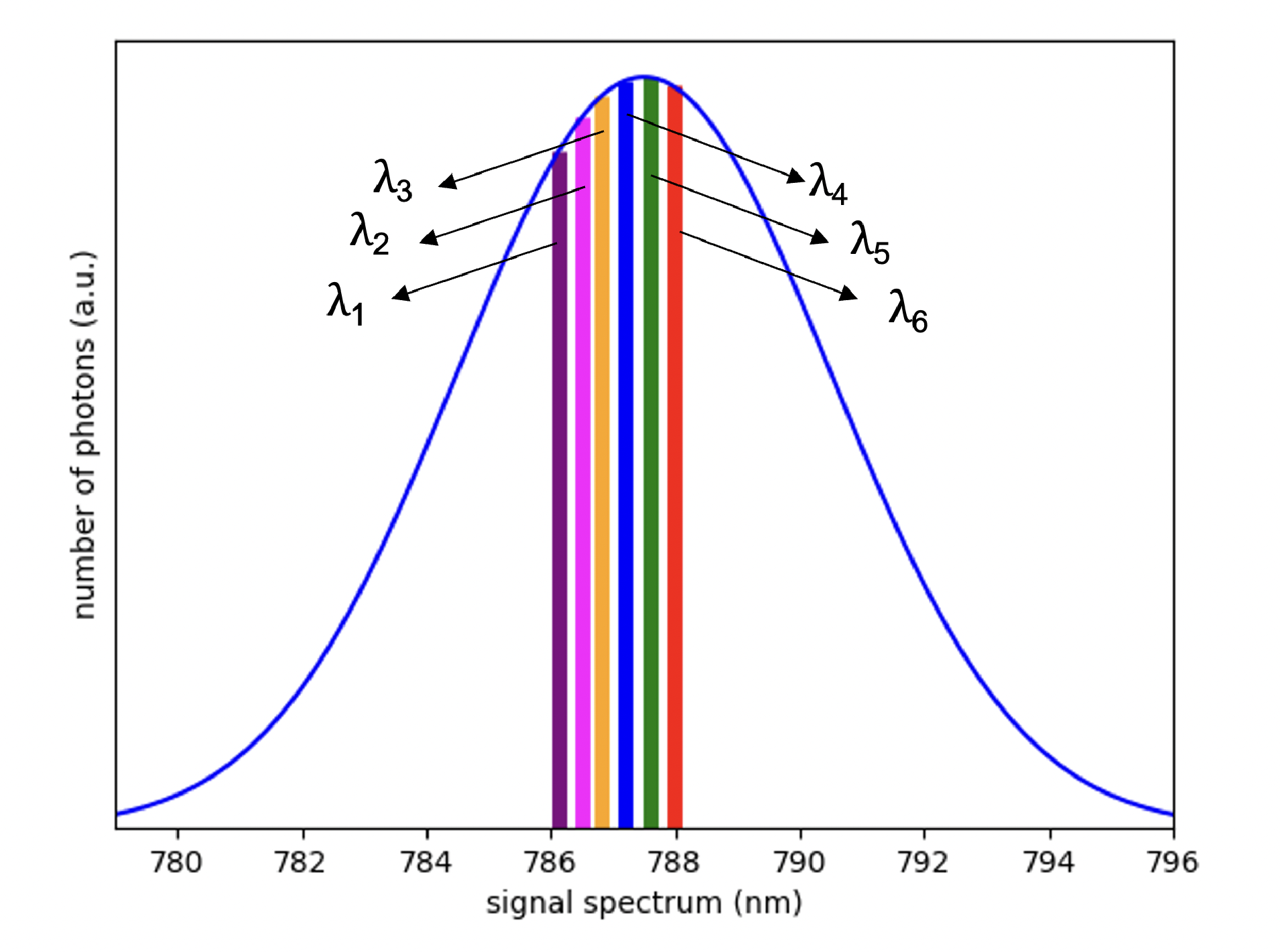}
    \caption{}
\end{subfigure}
\begin{subfigure}{0.49\textwidth}
    \includegraphics[width=0.94\textwidth]{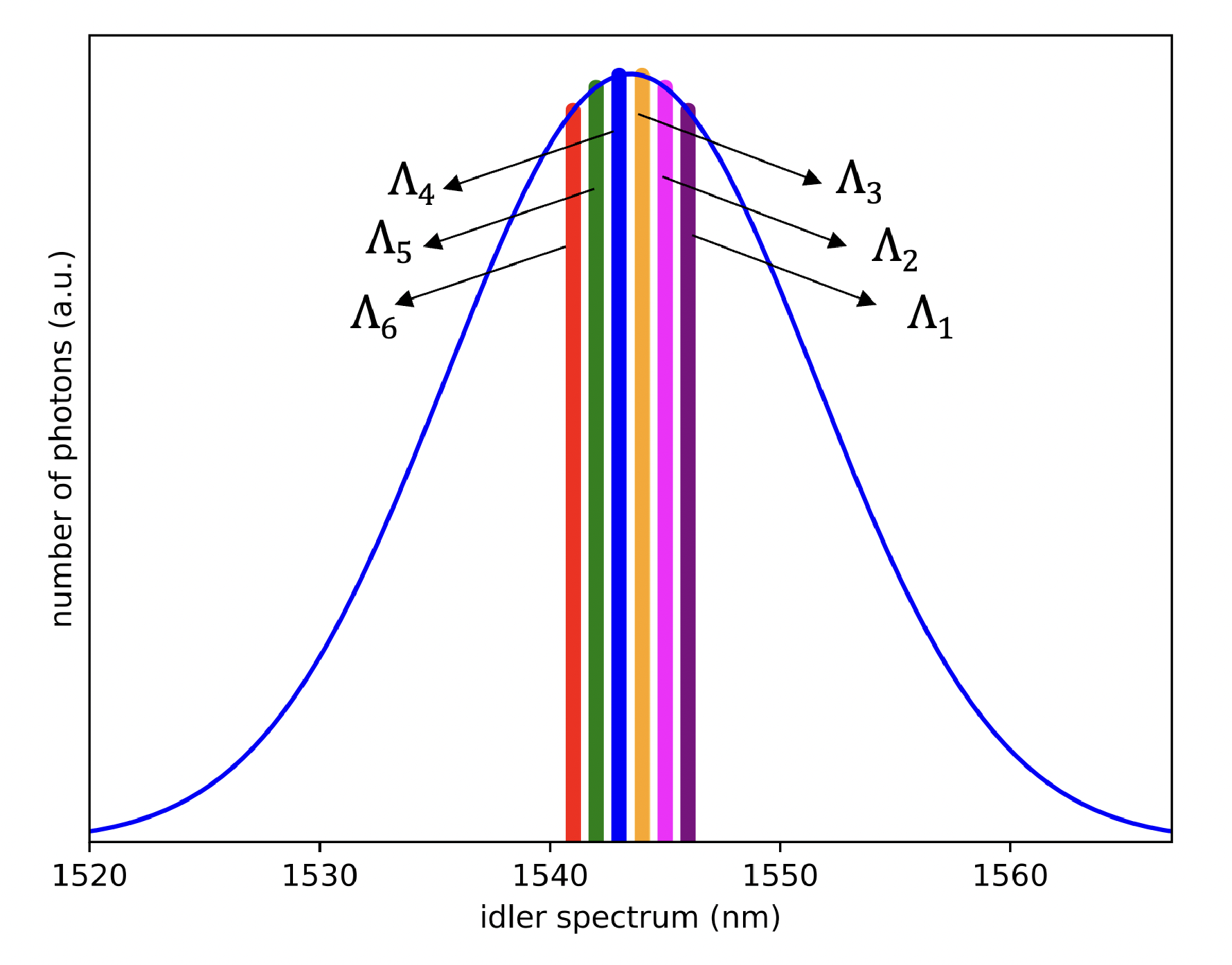}
    \caption{}
\end{subfigure}
\caption{The spectrum of the signal (a) and the idler (b) were calculated from the Sellmeier equations for PPMgOLN assuming a 2 nm pump bandwidth centered at 521.4 nm and a 15 mm crystal \cite{gayer_temperature_2008}. Wavelength correlated pairs $\{\lambda_{i},\Lambda_{i}\}$ are denoted by the same color in both spectrums and were chosen in accordance with Fig.~\ref{fig: alloc}. The idler channels match the ITU DWDM 100 GHz grid. } 
\label{spectrum}
\end{figure}

\begin{figure}
		\centering
		\includegraphics[width=0.5\textwidth]{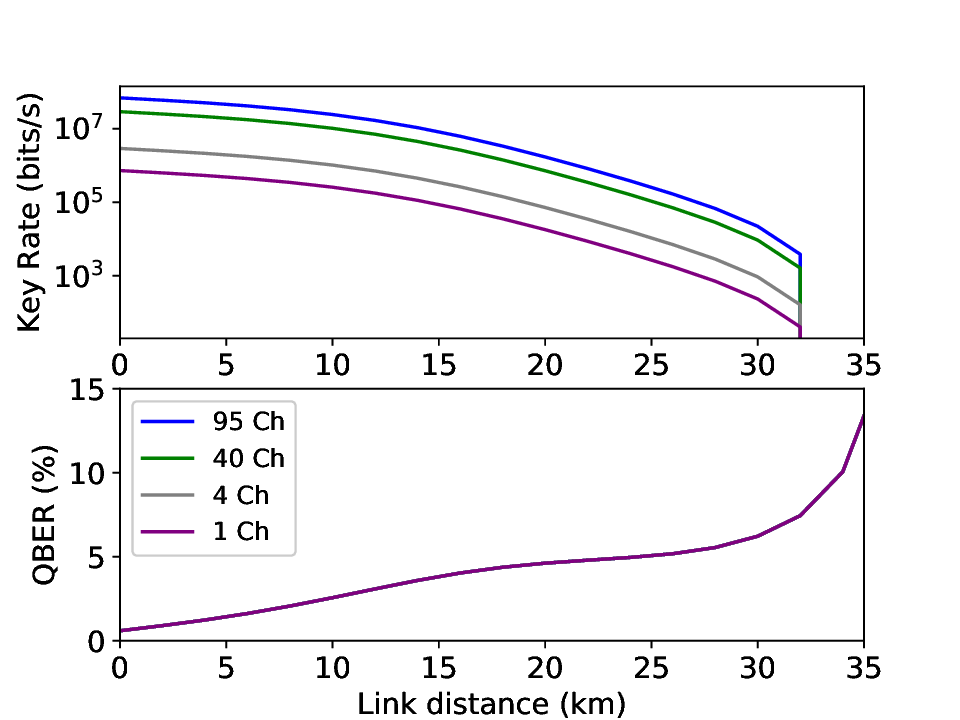}
	\caption{Scaling behaviour of the ground configuration ( Fig.~\ref{fig: alloc}c) for the QEYSSat compatible scenario considered in Table~\ref{tab:my_label} with the detector parameters given in Table~\ref{tab:param}.}
	\label{qeyssat}
\end{figure}

To further benchmark the potential of our network proposal, we analyze its performance for the QEYSSat mission use case. For this purpose, we assume the entangled photon source consists of a 15 mm long PPMgOLN waveguide pumped by a mode-locked pulsed laser centered at 521.4 nm with a bandwidth of 2 nm and a repetition rate of 80 MHz which produces polarization entangled photon pairs at 787.5 nm (signal) and 1543.2 nm (idler) via spontaneous parametric downconversion with respectively 15 nm and 39 nm of bandwidth. These assumptions on the entangled photon source are summarized in Table~\ref{tab:my_label}. 
\begin{table}[ht]
    \centering
      \captionsetup{justification=centering}
        \caption{Entangled photon source assumptions}
    \begin{tabular}{|c|c|}
    \hline
       Repetition rate  & 80 MHz \\
        Pump central wavelength   & 521.4 nm \\
        Pump bandwidth   & 2 nm \\
        Signal central wavelength   & 787.5 nm \\
        Signal bandwidth & 15 nm\\
        Idler central wavelength & 1543.2 nm\\
        Idler bandwidth & 39 nm\\    
        \hline
    \end{tabular}

    \label{tab:my_label}
\end{table}
The required wavelength channels for a hypothetical 4-node ground network and QEYSSat are illustrated in Fig.~\ref{spectrum}. The color coding is done in accordance with the channel allocation depicted in Fig.~\ref{fig: alloc}. The idler wavelength channels were chosen for compatibility with the standard International Telecommunication Union (ITU) 100 GHz grid. The corresponding correlated signal channel around $787.5$ nm was selected for low-loss transmission through the atmosphere and compability with mature high-efficiency Si-APDs \cite{Bourgoin_2013,IDQ}. Outside of the satellite pass, the 787.5 nm signal is rerouted to the ground users through standard telecom infrastructure. Whilst the telecom fibre is slightly multimode for the 787.5 nm light, Meyers-Scott et al. have shown that there is minimal crosstalk between the spatial modes over several kilometers resulting in a well-preserved polarization state and timing signature \cite{Meyers}. In spite of the greater fibre losses for the 787.5 nm light, the $ 32$ km range shown in Fig.~\ref{qeyssat} is large enough to span a metropolitan area network. As the ground user nodes are not resource-limited, we can circumvent the detector timing limitation via frequency demultiplexing to extend the linear performance increase to the low-loss regime. 
 Table~\ref{tab:month} shows the estimated key produced per month between ground users (ground configuration), and between a satellite and ground users (satellite configuration). Assuming that over a one-month period, one would typically have around 10 satellite passes with a typical 100 seconds quantum link duration, we estimate the total key generated monthly in both configurations by a user located 16 kilometres equidistant from all other users in the network and the OGS. In the ground configuration, the $95$ channels can connect up to 14 users with two preferred connections. The average key generated by a given user pair is then $\sim 1.9 \times 10^{11}$ bits. For the satellite configuration, the link loss will vary throughout the pass. We compute a conservative estimate by assuming an average link loss of $40$ dB for all 10 passes
which corresponds to a generated key of $\sim 7.5 \times 10^{7}$ bits. The proposed design presents a considerable scaling potential as the entire SPDC spectrum can be utilized. Assuming detector time resolutions as low as 35 ps \cite{35ps}, a $80$ MHz repetition rate and an ultra-dense WDM (UDWDM) with 12.5 GHz spacing, one could employ up to $\sim 350$ channels. The number of available frequency-time channels could be further increased by reducing the pulse repetition rate but at the expense of the single channel key rate.

\begin{table}
    \centering
      \captionsetup{justification=centering}
        \caption{Generated key rate over a one-month period for both network configurations}
    \begin{tabular}{|c|c|}
    \hline
       User-to-user ground distance  & 16 km \\
       Number of ground users in the network & 14\\
        User to OGS distance  &16 km \\
        Average number of passes & 10\\
        Average satellite pass duration  & 100 s\\
        Average satellite link loss  & 40~dB \\
        \hline
        Total key generated per ground user pair & $1.9 \times 10^{11}\hspace{0.1em}\rm bits$\\
        Total key generated between a ground node and satellite  & $7.5 \times 10^{7} \hspace{0.1em}\rm  bits $\\    
        \hline
    \end{tabular}
    \label{tab:month}
\end{table}

\section{Conclusion}\label{discussion}
We proposed and analyzed an  architecture suitable for serving both metropolitan networks or  ground-satellite links, reconfigurable with minimal  modifications. Our approach is uniquely designed to optimize the efficiency of satellite up-link scenarios such as QEYSSat, where the satellite receiver is only available for a few minutes every 24 hours. During the satellite pass, the secure key rate is improved through time and frequency multiplexing with no additional hardware requirements on the satellite. Additionally, the network's utility outside of the satellite pass is optimized by rerouting the signal to form a ground-based metropolitan network. The satellite can be exploited as a trusted node or entanglement swapping station to interconnect two metropolitan areas and extend the reach of quantum networks to a global scale. The scalability and simplicity render our network architecture a good candidate for commercial quantum communication networks.
\bmhead{Acknowledgements}
The authors would like to thank Duncan England for discussions on the frequency-to-time encoder.
\section*{Declarations}
\subsection*{Funding}
This work is supported by the  High Throughput and Secure Networks (HTSN) challenge by the National Research Council Canada, project HTSN-630. The authors further acknowlege   support from the Natural Sciences and Engineering Research Council of Canada (NSERC), the Canadian Foundation for Innovation (CFI), the Ontario Research Fund (ORF). SV would like to thank the NSERC PGS-D for personal funding.
\subsection*{Conflict of interest/Competing interests} 
The authors declare no conflicts of interest related to this article.
\subsection*{Code availability}
The simulation code used in this paper is available in the supplementary materials, Ref.~\cite{figshare}.
\begin{appendices}
\section{Quantum Key Distribution Simulation}\label{secA1}
The simulation code was written using the QuTIP toolbox in Python 
\cite{qutip,Holloway_paper}. We assume the protocol is symmetric with respect to the $X$ and $Z$ bases thus we find that $\delta_b=\delta_p$ and calculate the detection probabilities for the $Z$ basis according to the squashing model below with the following parameters:
\\\text{\hspace{2em}}$\chi$  \text{: squeezing parameter}
\\\text{\hspace{2em}loss\_a: Efficiency of the ground quantum channel}
\\\text{\hspace{2em}loss\_b: Efficiency of the satellite quantum channel}
\\\text{\hspace{2em}n\_factor\_a: Dark count rate in ground detectors}
\\\text{\hspace{2em}n\_factor\_b: Dark count rate in satellite detectors}
\\\text{\hspace{2em}dim: Fock space dimension}
\\\text{\hspace{2em}N: number of multiplexed channels}
\\\text{\hspace{2em}$\tau$: Detector dead time}
\\\text{\hspace{2em}f\_ec: error correction factor.}
\subsection{Squashing Model}
\begin{verbatim}
def measure_2folds_4modes_squashing_optimized(dim,chi,proj,proj2,tau):
    vacc = basis(dim,0)
    det_exp = zeros((2,2,2,2))
    H_sq = 1j*\chi*(tensor(create(dim),create(dim))+tensor(destroy(dim),destroy(dim)))
    U_sq= H_sq.expm()
    spdc =U_sq*tensor(vacc,vacc)
    psi_i = tensor(spdc,spdc)
    out_i = reshape(transpose(reshape(psi_i.full(), (dim,dim,dim,-1)),(0,3,2,1)),
    (dim*dim*dim*dim,-1))
    psi_i = Qobj(out_i,dims = [[dim, dim, dim, dim], [1, 1, 1, 1]])}
    for i in range(2):
       for j in range(2):
            for k in range(2):
               for l in range(2):
                 det_exp[i][j][k][l]} =
                 abs(expect(tensor(proj[i],proj[j],proj2[k],proj2[l]),psi_i))
    HH=(det_exp[0][1][0][1]+0.5*(det_exp[0][0][0][1]+det_exp[0][1][0][0])
    +0.25*det_exp[0][0][0][0])/(1+ tau*(det_exp[0][1][0][1]
    +0.5*(det_exp[0][0][0][1]+det_exp[0][1][0][0])+0.25*det_exp[0][0][0][0]))
    VV=(det_exp[1][0][1][0]+0.5*(det\_exp[0][0][1][0]+det_exp[1][0][0][0])
    +0.25*det_exp[0][0][0][0])/(1+\tau*(det_exp[1][0][1][0]
    +0.5*(det_exp[0][0][1][0]+det_exp[1][0][0][0])+0.25*det_exp[0][0][0][0]))
    HV=(det_exp[0][1][1][0]+0.5*(det_exp[0][0][1][0]+det_exp[0][1][0][0])
    +0.25*det_exp[0][0][0][0])/( 1+\tau*(det_exp[0][1][1][0]
    +0.5*(det_exp[0][0][1][0]+det_exp[0][1][0][0])+0.25*det_exp[0][0][0][0]))
    VH=(det_exp[1][0][0][1]+0.5*(det_exp[0][0][0][1]+det_exp[1][0][0][0])
    +0.25*det_exp[0][0][0][0])/(1+ \tau*(det_exp[1][0][0][1]
    +0.5*(det_exp[0][0][0][1]+det_exp[1][0][0][0])+0.25*det_exp[0][0][0][0]))
return [HH,HV,VH,VV]
\end{verbatim}
\subsection{Secure key rate calculation}
\begin{verbatim}
simqkdentanglement(chi,loss_a,loss_b,n_factor_a,n_factor_b,dim,N}):
    vacc= basis(N,0)
    qber_list=[]
    twofolds_list=[]
    for i in range(N):
        BucketDetector_realistic_detector(N,loss_a,n_factor_a)}
        b_det = BucketDetector_realistic_detector(N,loss_b,n_factor_b)}
        probs2f_i= measure_2folds_4modes_squashing_optimized(N,chi,a_det,b_det,tau)}
        twofolds_i=probs2f_i[0]+probs2f_i[1]+probs2f_i[2]+probs2f_i[3]
        twofolds_list.append(twofolds_i)}       
        qber_i=(probs2f_i[0]+probs2f_i[3])
        qber_list.append(qber_i)}
        twofolds=sum(twofolds_list)}
        qber=sum(qber_list)/sum(twofolds_list)}
        if qber>0:
            H2=-qber*log2(qber) - (1-qber)*log2(1-qber)}
        else:
            H2 = 0
    f_e = f_ec + qber
    skr=-1.0*real(twofolds*0.5*(1-(1+f_e)*H2))
return [skr,qber,twofolds]
\end{verbatim}
\end{appendices}

\end{document}